\documentclass[aps,prd,superscriptaddress]{revtex4}

\usepackage{amsfonts}
\usepackage{amsmath}
\usepackage{graphicx}
\usepackage{color}
\usepackage{verbatim}
\usepackage[dvips]{epsfig}
\usepackage{float}
\usepackage{epsfig,subfig}
\usepackage{epstopdf}
\usepackage{hyperref}
\usepackage{cleveref}

\DeclareMathAlphabet\mathbfcal{OMS}{cmsy}{b}{n}

\begin{document}

\title{Gravitational Searches for Lorentz Violation with Ultracold Neutrons}

\author{C. A. Escobar}
\email{carlos\_escobar@fisica.unam.mx}
\affiliation{Instituto de F\'{i}sica, Universidad Nacional Aut\'{o}noma de M\'{e}xico, Apdo. Postal 20-364, Ciudad de M\'{e}xico 01000, M\'exico}

\author{A. Mart\'{i}n-Ruiz}
\email{alberto.martin@nucleares.unam.mx}
\affiliation{Instituto de Ciencias Nucleares, Universidad Nacional Aut\'{o}noma de M\'{e}xico, 04510 Ciudad de M\'{e}xico, M\'{e}xico}
\affiliation{Centro de Ciencias de la Complejidad, Universidad Nacional Aut\'{o}noma de M\'{e}xico, 04510 Ciudad de M\'{e}xico, M\'{e}xico}

\begin{abstract}
We investigate the consequences of Lorentz violation (as expressed within the gravity sector of the Standard-Model Extension) for gravitational quantum states of ultracold neutrons (UCNs). Since our main aim is to compare our theoretical results with the recent high-sensitivity GRANIT experiment, we frame this work according to the laboratory conditions under which it was carried out. This offers the possibility of testing Lorentz invariance by experiments using UCNs. Thus we consider the nonrelativistic Hamiltonian describing the quantum mechanics of an unpolarized neutron's beam in presence of a weak-gravity field, and the latter is described by a post-Newtonian expansion of the metric up to order $O (2)$ and linear in the Lorentz-violating coefficients $\bar{s} ^{\mu \nu}$. Using a semi-classical wave packet, which is appropriate to describe an intense beam of UCNs, we derive the effective Hamiltonian describing the neutron's motion along the axis of free fall and then we compute the Lorentz-violating shifts on the energy levels. The comparison of our results with those obtained in the GRANIT experiment leads to an upper bound for a particular combination of the  Lorentz-violating coefficients.
\end{abstract}

\maketitle

\section{Introduction}

Deviations from Lorentz symmetry are predicted to occur in models of quantum gravity \cite{qg1}. The major difficulty behind a direct test of quantum gravity effects is the lack of experimentally accessible phenomena at the energy scale in which they operate ($\sim 10 ^{19}$GeV). Nevertheless, low-energy experiments provide an alternative route for an indirect test of quantum gravity. If violations of Lorentz symmetry are detected at low-energies, they will validate that a quantum theory of gravity should exist and may serve as a guiding principle for a fundamental theory. Within the context of effective field theories, a general Lorentz-violating (LV) framework, which includes the standard model of particle physics and general relativity, has been developed. This is known as the Standard-Model Extension (SME) \cite{SME1,SME2}. The SME provides an approach within which to analyze the results of experiments testing Lorentz violation. Therefore, currently there is a large experimental effort to find upper bounds on LV coefficients for the different sectors of the SME. Alternative scenarios with Lorentz violation have also been discussed in Refs. \cite{LV5, others2a}.

Here, we consider the physics of ultracold neutrons (UCNs) as a possible candidate to test Lorentz symmetry within the minimal-gravity SME sector (mgSME). To be precise, we examine the effects of the mgSME on the quantum free-fall of UCNs. In general, in the mgSME there are 20 coefficients for Lorentz violation organized into a scalar $u$, two-tensor $s ^{\mu \nu}$, and four-tensor $t ^{\mu \nu \alpha \beta}$, which are directly contracted with the Ricci scalar $R$, the trace-free Ricci tensor $R ^{T} _{\mu \nu}$ and the Weyl conformal tensor $C _{\mu \nu \alpha \beta}$, respectively. Moreover, to avoid inconsistencies with the Bianchi identity \cite{BIdent}, in curved space-times, Lorentz violation must arise spontaneously, and the dominant effects of the weak-field gravity are controlled by the vacuum expectation values $\bar{s} ^{\mu \nu}$ of the $s ^{\mu \nu}$-coefficients. To date, gravitational searches for Lorentz violation have included studies of gravitational waves \cite{GW}, cosmic rays \cite{CR}, gravimetry \cite{Gravimetry} and lunar laser ranging \cite{Lunar}.

Motivated by the recent high-sensitivity GRANIT experiment \cite{Granit}, and following the idea of our previous work \cite{AlbertoEscobar}, in this paper we aim to use the good precision achieved in such an experiment to set bounds on the LV coefficients of the mgSME. For a detailed description of the GRANIT experiment see Ref. \cite{Granit}. In short, they show that an intense beam of UCNs moving in Earth's gravity field does not bounce smoothly but at certain well-defined quantized heights, as predicted by quantum theory. In principle, both the fermion- and the gravity-sector of the SME affect the quantum mechanics of a freely falling neutron. Nevertheless, since current constraints on the LV coefficients $c _{\mu \nu}$ of the neutron sector are at the level of $10 ^{-8}$ to $10 ^{-29}$ \cite{datatables}, we can neglect them and focus our attention on the LV coefficients of the mgSME only. To be precise, our main goal is the calculation of the LV shifts on the energy levels of UCNs which, upon comparison with the maximal experimental precision achieved in the GRANIT experiment, will lead to an upper bound for the $\bar{s} ^{\mu \nu}$-coefficients of the mgSME. To this end, we frame this work according to the laboratory conditions under which experiments were carried out. The program we shall follow in this paper is the following.

To begin with, we consider the effective field theory which describes a single fermion in a curved spacetime background defined by the minimal-gravity SME sector. Since LV effects are expected to be small, we take the modified Einstein field equations in a post-Newtonian expansion of the metric up to order $O(2)$ and linear in the $\bar{s} ^{\mu \nu}$-coefficients. The matter-gravity coupling part of the action produces the corresponding Dirac Hamiltonian. Next, we work out the nonrelativistic expansion of the Hamiltonian and consider only the spin-independent terms, which are the ones relevant to describe the dynamics of an unpolarized beam of slow neutrons. Moreover, since the neutron's motion in the plane perpendicular to the axis of free fall (e.g. the $z$-axis) is governed by classical laws, we use a Gaussian wave-packet to derive an effective Hamiltonian which describes the effects of the mgSME on the UCNs. This program leads to a $z$- and $\hat{p} _{z}$-dependent effective potential which encodes the LV effects on the energy levels of the UCNs. Finally, using the formalism of nondegenerate perturbation theory we compute the LV energy shifts, which are the ones we compare with the results reported by the GRANIT experiment.

We begin in Sec. \ref{model1} by reviewing the formulation of the minimal gravity SME sector and we present the nonrelativistic Hamiltonian (including matter-gravity couplings) which describes spin-independent effects for a fermion in a LV weak gravitational field. Next, using a semi-classical wave-packet, in Sec. \ref{EffecHamil} we derive the effective Hamiltonian which describes the neutron's motion along the axis of free fall. The main calculations in the derivation of the effective LV potential are relegated to Appendix \ref{AEvalues}. Comparing the LV energy shifts and the experimental precision achieved in the GRANIT experiment, we set an upper bound for a combination of the LV coefficients in Sec. \ref{cotas1}. Finally in Sec. \ref{conclu1}, we summarize the main results of the paper. Throughout this work, we take the spacetime metric signature to be $(-,+,+,+)$.

\section{The model}
\label{model1}

Our main concern in this work is the description of the quantum mechanics of a single fermion in a Lorentz-violating weak gravitational field described by the mgSME. To this end, we summarize the action for the model, describe the post-Newtonian analysis of the linearized field equations for the pure gravity sector of the SME, and present the nonrelativistic Hamiltonian.

\subsection{Lorentz-violating gravity}

The effective action of the minimal-gravity SME sector (with vanishing torsion) locally coupled to a fermion field $\psi$ can be written as \cite{SignalPos}
\begin{align}
S = S _{\mbox{\scriptsize EH}} + S _{\mbox{\scriptsize LV}} + S _{\psi} . \label{action1}
\end{align}
The first term corresponds to the Einstein-Hilbert action of general relativity, which is given by
\begin{align}
S _{\mbox{\scriptsize EH}} = \frac{1}{2\kappa} \int e \left( R - 2 \Lambda \right) \, d ^{4} x , \label{actionEH}
\end{align}
where $R$ is the Ricci scalar, $\Lambda$ is the cosmological constant, $e$ the determinant of the vierbein and $\kappa = 8 \pi G c ^{-4}$. Since the focus of this work is the post-Newtonian limit of (\ref{action1}), in which the effects of $\Lambda$ are known to be negligible, we set $\Lambda=0$ in the rest of the work.

The second term in Eq. (\ref{action1}) contains the leading Lorentz-violating gravitational couplings. It can be written as
\begin{align}
S _{\mbox{\scriptsize LV}} = \frac{1}{2\kappa} \int  e \left( -u R + s ^{\mu\nu} R ^{T} _{\mu\nu} + t ^{\mu\nu\alpha\beta}C _{\mu\nu\alpha\beta} \right) d ^{4} x , \label{actionLV}
\end{align}
where $R ^{T} _{\mu \nu}$ is the trace-free Ricci tensor and $C _{\mu\nu\alpha\beta}$ is the Weyl conformal tensor. The LV coefficients $s ^{\mu\nu}$ and $t ^{\mu\nu\alpha\beta}$ inherit the symmetries of the Ricci tensor and Riemann curvature tensor, respectively. A total of 20 independent coefficients control the possible deviations of Lorentz symmetry. Since the $u$-coefficient is only a rescaling factor, it is unobservable and then it is discarded for this work \cite{SignalPos}. Because of a tensor identity, the 10 coefficients $t ^{\mu\nu\alpha\beta}$ vanish from the linearized equations \cite{SignalPos, YuriT}, and since we are ultimately focusing on the linearized theory, we can disregard them, thus leaving back the 9 coefficients $s ^{\mu \nu}$ in this limit. From now on, we will focus only on the $s ^{\mu\nu}$ coefficients. 

Finally, the action $S _{\psi}$ for a single fermion $\psi$ of mass $m$ is
\begin{align}
S _{\psi} = \int  e \left( \frac{1}{2}i  e^\mu\,_a \bar{\psi}\gamma^a\overleftrightarrow{D}_\mu\psi-m\bar{\psi}\psi \right) d^{4} x . 
\end{align}
In the present context, we consider the fermion action to be Lorentz symmetric. This is motivated by the fact that the $c _{\mu \nu}$-coefficients for the neutron sector of the SME, which are the ones relevant in the free fall of UCNs \cite{AlbertoEscobar}, have been strongly bounded by experiments, and thus we can focus our attention to the LV coefficients of the mgSME only.

\subsection{Post-Newtonian expansion}

The first step towards the analysis of the leading post-Newtonian corrections to general relativity induced by Lorentz violation is the linearization of the field equations. To avoid inconsistencies with the Bianchi identities of pseudo-Riemannian geometry \cite{BIdent}, it is  generally assumed that Lorentz violation must arise spontaneously in such a way that the LV coefficients are treated as dynamical fields that acquire nonzero vacuum expectation values. This means that the LV coefficients can be written as
\begin{align}
s ^{\mu \nu} = \bar{s} ^{\mu \nu} + \tilde{s} ^{\mu \nu} ,
\end{align}
where $\bar{s} ^{\mu \nu}$ and  $\tilde{s} ^{\mu \nu}$ denote the vacuum expectation values and the fluctuations of the LV coefficients, respectively. Note that the fluctuations could include massless Nambu-Goldstone modes and massive modes \cite{Modes1,Modes2}. Moreover, the vacuum values are taken as constants in a special observer coordinate system. Now, we can employ the usual linearization program of the field equations.

In linearized gravity the metric is expanded as
\begin{align}
g _{\mu\nu} &= \eta _{\mu\nu} + h _{\mu\nu} ,  \label{MetricExp}
\end{align}
where $\eta _{\mu \nu}$ is the Minkowski metric and $h _{\mu \nu}$ is the metric fluctuation. For a detailed derivation of the field equations as well as the linearized theory see Ref. \cite{SignalPos}. Here we just present the final results. Since the LV coefficients are small, for present purposes it suffices to work at linear order in the vacuum values $\bar{s}$, so in what follows nonlinear terms at $O(h ^{2})$, $O(h \tilde{s})$ and $O(\bar{s} ^{2})$, are disregarded. The resulting linearized field equations are
\begin{align}
G _{\mu \nu} = \kappa (T _{M}) _{\mu \nu} + \bar{s} ^{\kappa \lambda} R _{\kappa \mu \nu \lambda} - \bar{s} ^{\kappa} _{\phantom{\kappa} \mu} R _{\kappa \nu} - \bar{s} ^{\kappa} _{\phantom{\kappa} \nu} R _{\kappa \mu}  + \frac{1}{2}  \bar{s} _{\mu \nu} R + \eta _{\mu \nu}  \bar{s} ^{\kappa \lambda} R _{\kappa \lambda} ,  \label{linearizedFE}
\end{align}
where $R _{\mu \nu \alpha \beta}$ is the Riemann curvature tensor, $G _{\mu \nu}$ is the Einstein tensor, $R _{\mu \nu}$ is the Ricci tensor, and $R$ is the Ricci scalar. All theses tensors are understood as linearized in the metric fluctuation $h _{\mu \nu}$.  

Following standard techniques, we expand the linearized field equations (\ref{linearizedFE}) in a post-Newtonian series. As usual, the development of this approximation for the metric fluctuation requires the introduction of certain potentials for the perfect fluid. For the purposes of this work, and due to the smallness of the LV coefficients, it suffices to work at second order $O(2) = v \approx G m / r$, where $m$ is the typical body mass and $r$ is the typical system distance. For the pure-gravity sector of the minimal SME, the relevant potentials are
\begin{equation}
U = G\int  \frac{\rho(\vec{r} ^{\, \prime},t)}{R} d^3 \vec{r} ^{\, \prime} ,  \quad U^{ij} = G\int \frac{\rho(\vec{r} ^{\, \prime},t)R ^{i} R ^{j}}{R^3} d^3 \vec{r} ^{\, \prime}  ,  \label{potentials1a}
\end{equation}
where $\rho (\vec{r} , t)$ is the density of the mass distribution, $R ^{j} = \vec{r} ^{j} - \vec{r} ^{\, \prime j}$ and $R = \vert \vec{r}-\vec{r} ^{\, \prime} \vert$. Here, $U$ is the Newtonian gravitational potential, and $U ^{ij}$ lie beyond general relativity. Note that these potentials, related to the mass density, are the dominant terms in the Lorentz-violating post-Newtonian expansion. Higher-order corrections require additional potentials which are defined in terms of the mass current (all discussed in Ref. \cite{SignalPos}), however they lie beyond the scope of this work.

The components of the metric fluctuations can be obtained after an appropriate coordinate gauge choice. Here we impose the gauge conditions 
\begin{align}
\partial _{j} h _{0j} = \frac{1}{2} \partial _{0} h _{jj} , \qquad \partial _{j} h _{jk} = \frac{1}{2} \partial _{k} \left( h _{jj} - h _{00}  \right) , \label{GaugeCond}
\end{align}
which permit breaking the linearized field equations into its temporal and spatial components, and then expressing the metric $h _{\mu \nu}$ in terms of the potentials $U$ and $U ^{ij}$. To post-Newtonian $O(2)$ and linear order in the LV coefficients, the metric components $h _{\mu \nu}$ read
\begin{subequations}
\begin{align}
h _{00} &= \frac{1}{c ^{2}} \left[ (2 + 3 \bar{s} ^{00}) U  + \bar{s} ^{ij}  U ^{ij} \right] , \label{h00}  \\ h _{0j} &= - \frac{1}{c ^{2}} \left( \bar{s}^{0j} U - \bar{s}^{0k} U ^{jk} \right) , \label{h0j} \\  h _{ij} &= \frac{1}{c ^{2}} \left[ (2 - \bar{s} ^{00}) U + \bar{s} ^{lm} U ^{lm} \right] \delta ^{ij} - \frac{1}{c ^{2}} \left( \bar{s} ^{il} U ^{lj} + \bar{s} ^{jl} U ^{li} \right) + \frac{2}{c ^{2}} \bar{s} ^{00} U ^{ij} .  \label{hij}
\end{align}
\end{subequations}
As we shall see in the following, these metric perturbation components couple locally to the fermion fields, thus affecting the quantum mechanics of the particles.

\subsection{Quantum theory}

In order to investigate the effects of the minimal-gravity SME sector on the quantum mechanics of a nonrelativistic system we have to consider the associated Hamiltonian. The derivation is subtle because of the time-dependence arising from the Dirac equation obtained from $S _{\psi}$. The field-redefinition method, which has been systematically employed to construct Hamiltonians in the context of the SME, can also be employed in the mgSME case. Full details are given in Ref. \cite{KosteleckyM}. After an appropriate field-redefinition at the level of the action, the Dirac equation emerges with the conventional time-dependence, and thus the relativistic Hamiltonian can be properly identified. The Dirac Hamiltonian $H _{\mbox{\scriptsize D}}$ splits into pieces according to the perturbation order. The zeroth order, $H _{0}$, corresponds to the conventional Dirac Hamiltonian in Minkowski spacetime. The first-order correction, $H _{h}$, arises from the metric fluctuation $h _{\mu \nu}$. Higher-order corrections and those linear in the fermion SME coefficients are not of interest here.

In this work we are primarily interested in experiments with UCNs, in particular, the quantum states of neutrons in the gravitational field. Therefore, the nonrelativistic limit of the Dirac Hamiltonian $H _{\mbox{\scriptsize D}}$ is required. This can be obtained by employing the Foldy-Wouthuysen transformation, which is a systematic program to obtain the nonrelativistic content of relativistic Hamiltonians \cite{FW}. The needed Hamilton operator is only a particular case of the full nonrelativistic Hamiltonian derived in Ref. \cite{KosteleckyM}. To post-Newtonian $O(2)$ and linear order in the LV coefficients, the relevant spin-independent nonrelativistic Hamiltonian splits into two pieces. First, the conventional Minkowski-spacetime Hamiltonian, and second, the Lorentz-violating part (arising from the first-order Dirac Hamiltonian $H _{h}$):
\begin{equation}
H _{\mbox{\scriptsize NR}} = \frac{1}{2} m c ^{2} h _{00}  - h _{0k} p ^{k} c - \frac{1}{4m} h _{00} p ^{2} - \frac{1}{2m} h _{jk} p ^{j} p ^{k} . \label{H1a}
\end{equation}
In what follows, we focus on the effects of the Hamiltonian (\ref{H1a}) on the gravitational quantum states of UCNs.

\section{Effective Hamiltonian}
\label{EffecHamil}

In this section we derive the effective Hamiltonian $H _{z}$ describing the quantum free fall of UCNs in the gravitational field described by the post-Newtonian metric perturbations (\ref{h00})-(\ref{hij}). In the coordinate system attached to the Earth's surface [i.e. with $\vec{r} _{n} = (x,y,z)$ and $\vec{r} ^{\, \prime} = (0,0,- R _{\oplus})$ localizing the neutron and the Earth's center from the $z=0$ surface, respectively], the post-Newtonian potentials in Eq. (\ref{potentials1a}) can be written as
\begin{equation}
U = \frac{GM _{\oplus}}{r} , \qquad U ^{ij} = U \frac{r ^{i} r ^{j}}{r ^{2}} , \label{potentials1}
\end{equation}
where $G$ is the gravitational constant, $M _{\oplus}$ is the Earth's mass, $r = \sqrt{\vec{r}\cdot\vec{r}}$, and $\vec{r} \equiv \vec{r} _{n} - \vec{r} ^{\, \prime} = (x,y,z+R _{\oplus})$, being $R _{\oplus}$ the  Earth's radius. As usual, the effective quantum Hamiltonian can be obtained from the classical one (\ref{H1a}) by promoting the classical observables to quantum operators by means of the Weyl quantization rule. The Weyl rule associates to the product $q ^{n} p ^{m}$, between the classical coordinates and momenta, a quantum operator which is a totally symmetrized  linear combination of terms, each with $n$ factors of $q$ and $m$ factors of $p$ \cite{WeylQR}. The resulting Hamiltonian is given by
\begin{equation}
H _{\mbox{\scriptsize NR}} =  V _1+ V _2 + V _3 + V _4 ,
\end{equation}
where $V _{i}$, with $i=1,2,3,4$, corresponds to the quantum operator associated with each of the four terms in Eq. (\ref{H1a}). Explicitly we have
\begin{subequations}
\begin{align}
V _{1} &= \frac{1}{2}m c ^{2} h _{00} , \label{V1Pot}  \\ V _{2} &= - c \left( h _{0k} \hat{p} ^{k} + \frac{1}{2} h _{0k , k} \right) , \label{V2Pot} \\  V _{3} &= - \frac{1}{4m}\bigg( h _{00} \delta _{ij} \hat{p} ^{i} \hat{p} ^{j} + h_{00,i} \hat{p} ^{i} + \frac{1}{4} h _{00,ii} \bigg) , \label{V3Pot} \\ V _{4} &= -\frac{1}{2m} \bigg( h _{jk} \hat{p} ^{j} \hat{p} ^{k} +  h _{jk , j}  \hat{p} ^{k} + \frac{1}{4}  h _{jk,jk} \bigg) , \label{V4Pot}
\end{align}
\end{subequations}
where $\hat{p} _{i} = - i \hbar \partial _{i}$ is the momentum operator, $f _{,i} \equiv \hat{p} _{i} f$ and $f _{,ij} \equiv \hat{p} _{i} \hat{p} _{j} f$. Notice that $\hat{p} ^{i} = \hat{p} _{i}$ due to the chosen metric signature.

In what follows we work out the Hamiltonian to consider only the neutron vertical motion. To clarify the method let us review the GRANIT experiment performed at the Institute Laue-Langevin \cite{Granit}. They have found that UCNs freely falling in the Earth's gravity do not move continuously but jump from one height to another, such as quantum theory predicts. In the experiment,  they produce an intense horizontal beam of UCNs pointing slightly upwards and allow the neutrons to fall onto a horizontal mirror. By using a neutron absorber right above the mirror and counting the number of particles that move up to the absorber and down, they found that the heights of the neutrons are measured only at certain well defined values. In this situation, the vertical motion is quantized, while the horizontal one is driven by classical laws. According to the considerations above, it is valid to consider that the neutron's motion in the tangent plane to the Earth's surface, which is classical, can be described by a Gaussian wave packet of the form
\begin{align}
\psi ( \vec{r} _{\perp} ) = \frac{1}{\sqrt{\pi} \sigma} e ^{\frac{i}{\hbar}  \vec{p} _{\perp} \cdot  \vec{r} _{\perp} - \frac{ \vec{r} _{\perp} ^{2}}{2 \sigma ^{2}}} , 
\label{WPacket2}
\end{align}
where $\vec{r} _{\perp} = (x,y)$ and $ \vec{p} _{\perp} = (p _{x} , p _{y})$ are the coordinates and momentum in the plane perpendicular to the motion of free fall, respectively. A very small value of the characteristic width $\sigma$ of the wave packet assures that the classical condition is satisfied. The ansatz in Eq. (\ref{WPacket2}) allows us to derive a reduced one-dimensional Hamiltonian describing the neutron's vertical motion as follows
\begin{align}
H _{z} \equiv \left\langle H _{\mbox{\scriptsize NR}} \right\rangle = \int \psi ^{\ast} (  \vec{r} _{\perp}) \; H _{\mbox{\scriptsize NR}} \; \psi (  \vec{r} _{\perp}) d ^{2}  \vec{r} _{\perp} , \label{ReducedHamiltonian}
\end{align}
which indeed corresponds to the first order perturbation in the $x$-$y$ plane. Now we evaluate the expectation value of the quantum operators $V _{i}$, given by Eqs. (\ref{V1Pot})-(\ref{V4Pot}), in the semiclassical state (\ref{WPacket2}). The calculations are cumbersome, and thus we relegate the details to Appendix \ref{AEvalues}. We obtain $z$- and $\hat{p} _{z}$-dependent operators. Finally, we expand the result in a power series of $z$ and we consider only the leading order terms. We obtain
\begin{subequations}
\begin{align}
\langle V _{1} \rangle &= m U _{0} (1 + \gamma _{1} ) + m g z (1 +  \lambda _{1} ) - m g \frac{z ^{2}}{R _{\oplus}} , \label{V1-FIN} \\ \langle V _{2} \rangle &= m U _{0} \gamma _{2} + m g z \lambda _{2}  ,  \label{V2-FIN} \\ \langle V _{3} \rangle &= m U _{0} \gamma _{3} + m g z \lambda _{3}  - m g \frac{z ^{2}}{R _{\oplus}} \delta _{3} + \frac{\hat{p} ^{2} _{z} }{2 m} \xi _{3} + m g z \left( \frac{\hat{p} _{z}}{mc} \right) ^{2} \eta _{3} - \frac{1}{2} m g \frac{z ^{2}}{R _{\oplus}}  \left( \frac{\hat{p} _{z}}{mc} \right) ^{2},  \label{V3-FIN} \\ \langle V _{4} \rangle &= m U _{0} \gamma _{4} + m g z \lambda _{4} - m g \frac{z ^{2}}{R _{\oplus}} \delta  _{4}  + \frac{\hat{p} ^{2} _{z}}{2m} \xi _{4} + m g z  \left( \frac{\hat{p} _{z}}{mc} \right) ^{2} \eta _{4} - m g \frac{z ^{2}}{R _{\oplus}} \left( \frac{\hat{p} _{z}}{mc} \right) ^{2} ,  \label{V41-FIN}
\end{align}
\end{subequations}
where $U _{0} = - G M _{\oplus} / R _{\oplus}$ is the Newtonian potential on the Earth's surface and $g = G M _{\oplus} / R _{\oplus} ^{2}$ is the gravitational acceleration. In these expressions we have defined the $\bar{s} _{\mu \nu}$- and $\sigma$-dependent dimensionless quantities
\begin{subequations}
\label{definitions}
\begin{align}
\gamma _{1} & = \lambda _{1} + (\sigma / R _{\oplus} ) ^{2}  \equiv \frac{1}{2} (3 \bar{s} ^{00} + \bar{s} ^{zz}) - \frac{1}{2} (\sigma / R _{\oplus} ) ^{2} , \label{Gamma1}  \\ \gamma _{2} &= \lambda _{2} =  - \beta _{a} \bar{s} ^{0a}, \quad \beta _{a} = p _{a} / (mc) \\ \gamma _{3} &= \delta _{3} \gamma _{1}, \quad \lambda _{3} = \delta _{3} \lambda _{1}, \quad \xi _{3} = (U _{0} / c ^{2} ) \gamma _{1}, \quad \eta _{3} = \lambda _{1} / 2 , \quad \delta _{3} = \frac{1}{2} \left( \beta _{a} \beta _{a} + \Lambda ^{2} \right) , \quad \Lambda \equiv \hbar / (m c \sigma) \\ \gamma _{4} &= \lambda _{4}   - \delta _{3} (\sigma / R _{\oplus}) ^{2}, \quad \lambda _{4} = \delta _{3} \left( 2 - \bar{s} ^{aa} \right), \quad \delta _{4} = \delta _{3} (2 + \bar{s} ^{aa}) - \frac{1}{2} \beta _{b} \beta _{b} \bar{s} ^{aa}, \\ \xi _{4} &= ( U _{0} / c ^{2} )  \left[ 2 + \bar{s} ^{aa} - (\sigma / R _{\oplus}) ^{2} \right], \quad \eta _{4} = \frac{1}{2} \left[ 2 + \bar{s} ^{aa} - 3  (\sigma /R _{\oplus}) ^{2} \right] .
\end{align}
\end{subequations}

We observe that $\langle V _{1} \rangle$ consists of two terms. The first one is the unperturbed gravitational field up to second order in $z$, while the second one contains SME and quantum mechanical corrections. At the level of approximation we are considering, only the constant and linear terms receive additional corrections. Noticeably, $\langle V _{2} \rangle$ is a purely mgSME-potential. This is so because the nondiagonal metric components, $h _{0i}$, arise due to Lorentz symmetry breaking terms. To derive Eq. (\ref{V2-FIN}) we have used the fact that any quantity of the form $\left< F ( \vec{r} ) \right> \hat{p} ^{z}  + \frac{1}{2} \left< F ( \vec{r} ) \right> _{,z}$, for any smooth function $F ( \vec{r} )$, does not produce contributions to the energy shift. To demonstrate this result, we first note that it can be written as a total derivative term when acting on the $z$-dependent wave function $\chi _{n} (z)$,
\begin{align}
\hat{p} ^{z} \Big[ \chi _{n} (z) \left< F ( \vec{r} ) \right>  \chi _{n} (z) \Big]  = 2 \chi _{n} (z) \left[ \left< F ( \vec{r} ) \right> \hat{p} ^{z}  + \tfrac{1}{2} \left< F ( \vec{r} ) \right> _{,z} \right] \chi _{n} (z) , \label{def}
\end{align}
thus producing a boundary term when integrated over the physically allowed region $z \in [0, \infty )$, which is exactly zero due to the boundary conditions $\chi _{n} (z = 0) = 0$ and $\chi _{n} (z \rightarrow \infty) = 0$. In the derivation of Eqs. (\ref{V3-FIN}) and (\ref{V41-FIN}) we found many imaginary terms which fully cancel out due to the result of Eq. (\ref{def}). The details are presented in Appendix \ref{AEvalues}. 

Now we have the pieces to build up the reduced one dimensional Hamiltonian which describes the neutron's vertical motion. It can be conveniently expressed as 
\begin{equation}
H _{z} = \mathcal{U} _{0} +  H _{0} + V _{\sigma} + V _{\bar{s}} , \label{Hzfinal}
\end{equation}
where $\mathcal{U} _{0} = m U _{0} \left( 1 + \sum _{i = 1} ^{4} \gamma _{i} \right)$ collects constant terms, 
\begin{equation}
H _{0} = \frac{\hat{p} _{z} ^{2}}{2m} + mgz , \label{H_cero}
\end{equation}
is the conventional linearized Hamiltonian without Lorentz violation. The potentials
\begin{align}
 V _{\sigma} &= m g z \lambda _{\sigma} - m g \frac{z ^{2}}{R _{\oplus}} \delta _{\sigma}  + \frac{\hat{p} ^{2} _{z} }{2 m} \xi _{\sigma} + m g z \left( \frac{\hat{p} _{z}}{mc} \right) ^{2} \eta _{\sigma} - \frac{3}{2} m g \frac{z ^{2}}{R _{\oplus}}  \left( \frac{\hat{p} _{z}}{mc} \right) ^{2} ,  \label{Vsigma}
\end{align}
and
\begin{align}
 V _{\bar{s}} &= m g z \lambda _{\bar{s}} - m g \frac{z ^{2}}{R _{\oplus}} \delta _{\bar{s}}  + \frac{\hat{p} ^{2} _{z} }{2 m} \xi _{\bar{s}} + m g z \left( \frac{\hat{p} _{z}}{mc} \right) ^{2} \eta  _{\bar{s}} ,  \label{Vs}
\end{align}
contain all the corrections from the metric fluctuations and the LV coefficients, respectively. From Eq. (\ref{definitions}) we determine the parameters as
\begin{subequations}
\begin{align}
\lambda _{\sigma} &= 2 \delta _{3} - \frac{3}{2} (\sigma /R _{\oplus}) ^{2} (1 + \delta _{3}) , \quad \lambda _{\bar{s}} = \frac{1}{2} (3 \bar{s} ^{00} + \bar{s} ^{zz}) + \frac{1}{2} \delta _{3} (\bar{s} ^{00} + 3 \bar{s} ^{zz}) - \beta _{a} \bar{s} ^{0a} , \\ \delta _{\sigma} &= 1 + 3 \delta _{3} , \quad \delta _{\bar{s}} = (\Lambda ^{2} / 2) \bar{s} ^{aa} , \quad \xi _{\sigma} = ( U _{0} / 2c ^{2} )  \left[ 4 - 3 (\sigma /R _{\oplus}) ^{2} \right] , \quad \xi _{\bar{s}} = ( U _{0} / 2c ^{2} ) (5 \bar{s} ^{00} - \bar{s} ^{zz}) , \\ \eta _{\sigma} &= 1 - (9/4) (\sigma /R _{\oplus}) ^{2} , \quad \eta _{\bar{s}} = (5 \bar{s} ^{00} - \bar{s} ^{zz}) / 4 .
\end{align}
\end{subequations}
In the next section we analyze the influence of Lorentz-violating terms on the nonrelativistic energy levels of UCNs in the Earth's gravity.

\section{Energy Shifts and bounds on the $\bar{s} ^{\mu\nu}$ SME coefficients}
\label{cotas1}

The first term in Eq. (\ref{Hzfinal}), $\mathcal{U} _{0}$, does not imply any physical change on the neutron's energy spectrum; whereas the second term, $H _{0}$, is the well-known Hamiltonian describing the stationary energy eigenstates $\chi _{n} (z)$ of the UCNs in the Earth's gravity. Explicitly we have
\begin{align}
\chi _{n} (z) = \frac{1}{\sqrt{l _{0}}} \frac{\mbox{Ai} (a _{n} + z/l _{0} )}{\mbox{Ai} ^{\prime} (a _{n})} \Theta (z) , \label{UnpWaveFunc}
\end{align}
where $a _{n}$ is the $n$-th zero of the Airy function $\mbox{Ai}$, $l _{0} = \sqrt[3]{\hbar ^{2} / (2m ^{2}g ) }$ is the gravitational length and $\Theta (z)$ is the Heaviside function. The quantum state energies, defined by the boundary condition at $z=0$, are given by $E _{n} = - mgl_{0} a _{n} $. The height of a neutron with energy $E _{n}$ in the gravitational field, within the classical description, is found to be $h _{n} = E _{n}/(mg) = - a _{n} l _{0}$. 

To evaluate the alterations yielded by the potentials $V _{\sigma}$ and $V _{\bar{s}}$, the energy corrections are worked out as a first order perturbation on the corresponding neutron's wave functions $\chi _{n}$, that is,
\begin{equation}
\Delta E _{n} = \int \chi _{n} ^{\ast} (V _{\sigma} + V _{\bar{s}}) \chi _{n} dz .
\end{equation}
Using the properties of the Airy functions \cite{Airys,Airys2}, one can derive the following results
\begin{equation}
\begin{aligned}
 mg \left\langle z \right\rangle = \frac{2}{3} E _{n} , \quad \left\langle \hat{p} _{z} ^{2} / 2m \right\rangle = \frac{1}{3} E _{n},\quad  \frac{g}{2mc ^{2}} \left\langle z \hat{p} _{z} ^{2} \right\rangle = - \frac{2}{15} a _{n} E _{n} \frac{g l _{0}}{c ^{2}}  , \\ mg \langle z^2\rangle=\frac{8}{15}\frac{E_n^2}{mg}, \quad\quad \frac{\langle z^2 \hat{p}_z^2\rangle}{2m}=\frac{8}{105}\frac{E_n^3}{m^2g^2}-\frac{3}{7}mg l_0^3 ,
\end{aligned}
\label{ExpValues}
\end{equation}
which yield the energy shifts
\begin{equation}
\frac{\Delta E _{n}}{E _{n}}=\frac{2}{3}( \lambda _{\sigma} + \lambda _{\bar{s}} ) -\frac{8}{15}\frac{E_n}{mgR _{\oplus}} (  \delta _{\sigma} + \delta _{\bar{s}} )+\frac{1}{3} ( \xi _{\sigma} + \xi _{\bar{s}} ) +\frac{4}{15}\frac{E_n}{mc^2} ( \eta _{\sigma} + \eta _{\bar{s}} ) -\frac{8}{35}\frac{E_n^2}{(mc^2)(mgR _{\oplus})}, \label{EnergyShift}
\end{equation}
where the parameters $\lambda _{\sigma}$, $\delta _{\sigma}$, $\xi _{\sigma}$ and $\eta _{\sigma}$ collect the contributions from the metric fluctuations, and those parameters with the subscript $\bar{s}$ comprise the contributions from the coefficients for Lorentz violation. This energy correction $\Delta E _{n} $, together with the maximal experimental uncertainty in the GRANIT experiment $\Delta E _{n} ^{\scriptsize \mbox{exp}}$, may be used to set up an upper bound on the magnitude of the LV coefficients.

In spite of the weakness of the gravitational interaction and the number of systematic errors in laboratory conditions, the GRANIT experiment has recently confirmed this quantum-mechanical behavior where a noncoherent beam of UCNs propagating upwards in the Earth's gravity field produces quantized heights only. The values of the two lowest experimental heights are \cite{Nesvizhevsky3}
\begin{equation}
\begin{aligned}
 h _{1} ^{\mbox{\scriptsize exp}} &= \left( 12.2 \pm 1.8 _{\scriptsize \mbox{sys}} \pm 0.7 _{\scriptsize \mbox{stat}} \right) \mu \mbox{m} , \\ h _{2} ^{\mbox{\scriptsize exp}} &= \left( 21.6 \pm 2.2 _{\scriptsize \mbox{sys}} \pm 0.7 _{\scriptsize \mbox{stat}} \right) \mu \mbox{m} ,
\end{aligned}
\label{HeightsExp}
\end{equation}
such that the theoretical values, $h _{1} = 13.7 \mu \mbox{m}$ and $h _{2} = 24.0 \mu \mbox{m}$ \cite{Granit}, are therefore settled within the error bars. The good agreement between experiment and theory has been used for exploring deviations from the standard theory due to an eventual new physical mechanism, for example, to constrain axion-like interactions \cite{Axion}, short-range gravitational interactions \cite{ShortRange}  and the fundamental length scale in polymer quantum mechanics \cite{Martin}. Below, we will use the aforementioned results to set up an upper bound on the $\bar{s} ^{\mu \nu}$-coefficients.

Under the conditions in which the GRANIT experiment is performed, the nonrelativistic neutrons in low quantum states satisfy: $U _{0} / c ^{2} \approx 10 ^{-10}$, $E _{n} / (mc ^{2}) \approx 10 ^{-22}$, $E _{n} / (mg R _{\oplus}) \approx 10 ^{-31}$,  $\beta _{a} \approx 10 ^{-7}$, $\Lambda ^{2}  \approx 10 ^{-15}$ and $(\sigma / R _{\oplus}) ^{2} \approx 10 ^{-30}$. Clearly, most of the terms appearing in the energy shift (\ref{EnergyShift}) will be strongly suppressed. Assuming that the corrections produced by the LV coefficients are greater than those produced by the Lorentz-symmetric perturbations, we set an upper bound to the $\bar{s} ^{\mu \nu}$ coefficients. The experimental data for the first two lowest quantum states provide the values $\vert \Delta E _{1} ^{\textrm{exp}} \vert = 0.102 \textrm{peV}$ and $\vert \Delta E_2^{\textrm{exp}} \vert = 0.051 \textrm{peV}$ \cite{Nesvizhevsky3}. By imposing $\vert \Delta E _{n} \vert < \vert \Delta E _{n} ^{\textrm{exp}} \vert$ we find
\begin{equation}
\vert 3 \bar{s} ^{00} + \bar{s} ^{zz} \vert < 10 ^{-2} . \label{cota2}
\end{equation}
The energy shift (\ref{EnergyShift}) is given in the laboratory frame, where the LV coefficients take the constant values $s _{\mu \nu}$, with $\mu , \nu = t,x,y,z$. However, the laboratory frame rotates with the Earth, so the spatial components of $s _{\mu \nu}$ oscillate periodically as functions of the sidereal time $t$. Then, it is important to work in an appropriate inertial frame, such as the Sun-centered celestial-equatorial frame, which is  effectively inertial over the time scale of most Earth-based experiments. This induces corresponding variations in the observed energy shifts, with periodicities controlled by the Earth's sidereal rotation frequency $\Omega = 2 \pi / (23 \, \mbox{h} \, 56 \, \mbox{m} )$. In the Sun-centered frame the LV coefficients are given by $s _{\mu \nu}$, where $\mu , \nu = T, X, Y, Z$. For a detailed review of the transformation between the laboratory and the Sun-centered frames see Ref. \cite{SignalPos}. Data in the GRANIT experiment are usually taken at different sidereal times; therefore the energy shifts will only be sensitive to the time-averaged effect of the Lorentz-violating terms. The final result for the time-averaged combination appearing in Eq. (\ref{cota2}) yields
\begin{equation}
\vert 3 \bar{s} ^{TT} + 0.24 \left( \bar{s} ^{XX} + \bar{s} ^{YY} - 2.16 \bar{s} ^{ZZ} \right) \vert < 10 ^{-2} , \label{cota3}
\end{equation}
 where we have taken the Grenoble's colatitude as $\chi \approx 44.83 ^{\circ}$. Certainly, this bound is far from the expected values for the SME coefficients but it can be improved with future improvements in the experimental precision. 

We close this section indicating some of the reported current bounds on the minimal gravity SME sector. For example, gravitational waves set the bound $\bar{s}^{(4)}_{00}<10^{-15}$ \cite{GW}. Similarly, gravimetry tests provide the bounds $\bar{s}^{XX}-\bar{s}^{YY},\bar{s}^{XY}<10^{-9}$ and $\bar{s}^{XZ},\bar{s}^{YZ}<10^{-10}$ \cite{Gravimetry}. Other bounds coming from binary pulsars establish $\bar{s}^{XX}+\bar{s}^{YY}-2\bar{s}^{ZZ}<10^{-11}$ \cite{pulsar}, which in combination with our result in Eq. (\ref{cota3}) can be understood as a bound on the single coefficient $\vert \bar{s} ^{TT} \vert < 10 ^{-3}$, which competes with the one imposed from the Gravity Probe B \cite{GPB}. As expected, astronomic measurements or experiments at high energies lead to the tightest bounds for Lorentz violation; however, we cannot abandon other scenarios since they can guide us to new bounds for specific combinations for the LV coefficients, as is the case in the present work.

\section{Conclusion}
\label{conclu1}

Motivated by the recent high-sensitivity GRANIT experiments, in this paper we have investigated the effects of the minimal gravity sector of the Standard-Model Extension (SME) upon the gravitational quantum states of ultracold neutrons (UCNs). In other words, we have considered the physics of UCNs as a test bed for studying deviations from Lorentz symmetry at low energies.

In short, the Grenoble's group has shown that an intense beam of UCNs moving in the Earth's gravity field does not bounce smoothly but at certain well-defined quantized heights, which indeed is a direct consequence of the discrete energy levels of the system (by means of the identification $h _{n} =  E _{n} / mg$). Therefore, due to the good precision achieved in such experiments, any deviation from the quantum-mechanical prediction (due to an eventual new physical mechanism) can be tested. This idea has been used to set bounds on the coupling of short-range gravitational interactions and the fundamental length in polymer quantum physics. In view of this, in this paper we have investigated how the minimal gravity SME sector affects the energy levels of UCNs. To this end, we start with a Lorentz-violating extension of Einstein's general relativity and analyze the quantum free fall of UCNs in the proximity of the Earth's surface. Since we aim to compare our theoretical results with the quantized heights measured in the GRANIT experiments, we have framed this work according to the laboratory conditions under which experiments were carried out.

The first step was the calculation of the gravity field near the Earth's surface. Since Lorentz violation has not been detected yet in experiments, it is generally assumed that LV coefficients have small components in Earth-based laboratories, thus leading to very tiny modifications in physically measurable quantities. This means that we can solve the modified Einstein field equations in a post-Newtonian expansion of the metric. Due to the smallness of the $\bar{s} ^{\mu \nu}$-coefficients, here we only take into account linear-order terms in $\bar{s} ^{\mu \nu}$ and the post-Newtonian expansion up to second order. Next, once we know the behavior of the local gravity field, we need to consider the quantum-mechanics of a single fermion moving under its influence. To this end, we take the associated relativistic Dirac Hamiltonian and then compute its nonrelativistic limit by means of the Foldy-Wouthuysen procedure, which is appropriate to describe unpolarized UCNs. Given that the horizontal motion of UCNs is governed by classical laws, we use a semi-classical wave packet to obtain an effective Hamiltonian $H _{z} (z, \hat{p} _{z})$ describing the neutrons motion along the axis of free fall. We have worked out the energy shifts to first order in perturbation theory, and we found that they contain SME-, quantum- and relativistic-corrections. A comparison with the current experimental precision in the GRANIT experiment produces an upper bound of the order of $10^{-2}$ for the combination $\vert 3 \bar{s} ^{00} + \bar{s} ^{zz} \vert$ in the laboratory frame, which can be expressed as $\vert 3 \bar{s} ^{TT} + 0.24 \left( \bar{s} ^{XX} + \bar{s} ^{YY} - 2.16 \bar{s} ^{ZZ} \right) \vert $ in the Sun-centered frame.

Finally, we comment on three additional experiments involving UCNs which can also be used to enhance the bounds on the LV-coefficients: gravity-resonance-spectroscopy \cite{Jenke}, acoustic Rabi oscillations \cite{abele} and the neutron whispering gallery wave \cite{centrifugal}. The former experiment is based on the measurement of transitions induced by means of a mechanical oscillation of mirrors. In this case, the experimental precision is of the order of $10 ^{-14}$eV, which implies that the bound on the LV-coefficients can be improved by one order of magnitude as compared with the one reported here. We also mention the recent measurements done by the qBOUNCE collaboration \cite{abele}, which has measured the energy levels of UCNs with a precision of $10 ^{-15}$eV, resulting in much more precise limits for the LV-coefficients. The latter experiment deals with the long-living centrifugal quantum states of UCNs scattered on a cylindrical surface. In this case, the neutrons move in a fictitious gravity field of strength $10 ^{5}-10 ^{7}\,g$ at energies of the order of neV, which together with the current experimental precision could improve our upper bound by five orders of magnitude. We leave these systems for future investigations.

\acknowledgments

C. A. E. is supported by UNAM-DGAPA postdoctoral fellowship and the project PAPIIT No. IN111518. We greatly appreciate correspondence with Professor V. V. Nesvizhevsky.

\appendix

\section{Expectation values}
\label{AEvalues}

In this section we compute the expectation values of the quantum operators $V _{i}$ in the semi-classical state (\ref{WPacket2}). From now on, Latin indices of the middle of the alphabet $(i,j,k,l)$ refer to the three spatial components $x,y,z$; while Latin indices from the beginning of the alphabet $(a,b,c,e)$ refer to the coordinates $x,y$.

Let us first compute $\langle V _{1} \rangle$. In order to disentangle the effects on the neutron's vertical motion, it's convenient to split $\bar{s} ^{ij}U ^{ij}$ into their vertical ($z$-axis) and perpendicular ($xy$-plane) parts, namely $\bar{s} ^{ij}U ^{ij} = \bar{s} ^{ab}U ^{ab} + 2 \bar{s} ^{az}U ^{az} + \bar{s} ^{zz}U ^{zz}$. Using the result $\langle U ^{az} \rangle = 0$ (which follows from the axial symmetry around the axis of free fall), Eq. (\ref{V1Pot}) simplifies to $\langle V _{1} \rangle= - \frac{1}{2} m  \left[ (2 + 3 \bar{s} ^{00}) \langle U \rangle + \langle \bar{s} ^{ab} U ^{ab} \rangle + \langle \bar{s} ^{zz} U ^{zz} \rangle \right]$. The axial symmetry again implies that $\langle U ^{ab} \rangle = (\delta _{ab} / 2) \langle U \rho ^{2} / r ^{2} \rangle$ and from Eq. (\ref{potentials1}) we find the identity $\langle U \rangle  = \langle U \rho ^{2} / r ^{2} \rangle  + \langle U ^{zz} \rangle$. These results yield
\begin{align}
\left< V _{1} \right> &= - \frac{1}{2} m \left[ (2 + 3 \bar{s} ^{00} + \bar{s} ^{zz}) \left< U \right> - \left( \bar{s} ^{zz}  - \bar{s} ^{aa} / 2 \right) \left< U \frac{\rho ^{2}}{r ^{2}} \right> \right] . \label{V1-2}
\end{align}
The required expectation values in the state (\ref{WPacket2}) can be computed in a simple fashion. The result is
 \begin{align}
  \left< U \right> &= \frac{GM _{\oplus}}{\sigma} \sqrt{\pi} e ^{\xi ^{2}} \mbox{erfc} (\xi)  , \label{U} \\  \left< U \frac{\rho ^{2}}{r ^{2}} \right> &= \frac{GM _{\oplus}}{\sigma} \left[ (2 \xi ^{2} +1) \sqrt{\pi} e^{\xi ^{2}} \mbox{erfc} (\xi) - 2 \xi \right] , \label{U(rho/r)2}
\end{align}
where $\mbox{erfc} (\xi)$ is the complementary error function and $\xi \equiv (R _{\oplus} + z) / \sigma$. In practice, the experiments with UCNs bouncing on a horizontal mirror are very localized as compared with the Earth's radius, and thus we may approximate the expectation values in Eqs. (\ref{U}) and (\ref{U(rho/r)2}) for $R _{\oplus} \gg z$ and $R _{\oplus} \gg \sigma$. Using the asymptotic expansion of the complementary error function for large real arguments we can easily compute the asymptotic behavior of Eqs. (\ref{U}) and (\ref{U(rho/r)2}). Inserting the leading orders into Eq. (\ref{V1-2}), after some algebra we finally establish Eq. (\ref{V1-FIN}).

Now let's focus on $\langle V _{2} \rangle$. Decomposing $V _{2}$ into its vertical ($z$) and perpendicular ($x,y$) components, Eq. (\ref{V2Pot}) simplifies to $\left< V _{2} \right> = - c \left( \left< h _{0a} \hat{p} ^{a} \right> + \frac{1}{2} \left< h _{0a,a} \right> \right) - c \left( \left< h _{0z} \right>  \hat{p} ^{z} + \frac{1}{2} \left< h _{0z,z} \right> \right)$. As discussed in the main text, the second term does not produce corrections to the energy levels since it is a boundary term which we drop in what follows. Then we are left only with the first term. Substituting the metric fluctuation and disregarding those terms which vanish due to the axial symmetry we obtain
\begin{align}
\left< V _{2} \right> = \frac{1}{c} \bar{s} ^{0a} \left(  \left< U \hat{p} ^{a} \right> + \left< U ^{ab} \hat{p} ^{b} \right> \right) + \frac{1}{c} \bar{s} ^{0z}  \left( \left< U ^{az} \hat{p} ^{a} \right> + \frac{1}{2} \left< U ^{az} _{\phantom{az},a} \right> \right)  .  \label{V2-5}
\end{align}
Now we evaluate separately each of the required expectation values. Since $\left< U \hat{p} ^{a} \right> = p _{a} \left< U \right>$, $\left< U ^{ba} \hat{p} ^{b} \right> = p ^{b} \left< U ^{ba} \right>$,
$\left< U ^{az} \hat{p} ^{a} \right> = i \hbar (z +R _{\oplus}) \left< \frac{\rho ^{2} U}{\sigma ^{2} r ^{2}} \right>$ and $\left< \frac{1}{2} U ^{az} _{\phantom{az} ,a} \right> = - i \hbar (z+R _{\oplus}) \left< \frac{U}{r ^{2}} - \frac{3 \rho ^{2} U}{2r ^{4}}  \right>$, then Eq. (\ref{V2-5}) becomes
\begin{align}
\left< V _{2} \right> = \frac{1}{c} \bar{s} ^{0a}  p _{a} \left[ \left< U \right> + \left< U \frac{\rho ^{2}}{2r ^{2}}  \right> \right] + \frac{i \hbar}{c} \bar{s} ^{0z} (z+R _{\oplus}) \left< \frac{\rho ^{2} U}{\sigma ^{2} r ^{2}} - \frac{U}{r ^{2}} + \frac{3 \rho ^{2} U}{2r ^{4}}  \right> .  \label{V2-6}
\end{align}
In addition to the expectation values (\ref{U}) and (\ref{U(rho/r)2}), in this expression we also need the following results:
\begin{align}
\left< \frac{U}{r ^{2}} \right> &= \frac{2 GM _{\oplus}}{\sigma ^{3}} \left[ \frac{1}{\xi} - \sqrt{\pi} e^{\xi ^{2}} \mbox{erfc} (\xi) \right] , \\ \left< U \frac{\rho ^{2}}{r ^{4}}  \right> &=  \frac{2GM _{\oplus}}{3 \sigma ^{3}} \left[ \frac{2 (1 + \xi ^{2})}{\xi} - (2 \xi ^{2} + 3) \sqrt{\pi} e^{\xi ^{2}} \mbox{erfc} (\xi) \right] ,
\end{align}
from which we can directly verify that the complex term appearing in (\ref{V2-6}) vanishes. Thus we are left only with the first term in (\ref{V2-6}). Finally, inserting the asymptotic expansions of Eqs. (\ref{U}) and (\ref{U(rho/r)2}) into the first term in (\ref{V2-6}) we obtain Eq. (\ref{V2-FIN}). 

The calculations needed for the derivation of $\left< V _{3} \right>$ are more intricate than those required for the previous terms. Here we describe the main steps and present the final results. We start by decomposing $V _{3}$ into its perpendicular ($x,y$) and vertical ($z$) components, namely, $\langle V _{3} \rangle ^{(1)} = - \frac{1}{4 m} \left( \left< h _{00} \hat{p} ^{a} \hat{p} ^{a} \right> + \left< h_{00,a} \hat{p} ^{a} \right> + \frac{1}{4} \left< h _{00,aa} \right> \right) \equiv \left< V _{3} \right> ^{(1,1)} + \left< V _{3} \right> ^{(1,2)} + \left< V _{3} \right> ^{(1,3)}$ and $\langle V _{3} \rangle ^{(2)} = - \frac{1}{4 m} \left( \left< h _{00} \right> \hat{p} ^{2} _{z} + \left< h_{00} \right> _{, z} \hat{p} ^{z} + \frac{1}{4} \left< h _{00} \right> _{,zz}  \right)$. Inserting the metric perturbation $h _{00}$ and taking the required derivatives we can write each term in $\langle V _{3} \rangle ^{(1)}$ as
\begin{subequations}
\begin{align}
\left< V _{3} \right> ^{(1,1)} &= - \frac{1}{4m c ^{2}} \left( p _{a} p _{a} + \frac{2 \hbar ^{2}}{\sigma ^{2}}  \right) \left[ (2 + 3 \bar{s} ^{00}) \left< U \right> + \frac{1}{2} \bar{s} ^{aa}  \left< U \frac{\rho ^{2}}{r ^{2}} \right> + \bar{s} ^{zz} (z + R _{\oplus} ) ^{2} \left<  \frac{U}{r ^{2}} \right> \right] \notag \\ & \phantom{=} + \frac{\hbar ^{2}}{4m c ^{2} \sigma ^{4}} \left[ ( 2 + 3 \bar{s} ^{00} ) \left< \rho ^{2} U \right> + \frac{1}{2} \bar{s} ^{aa} \left< U \frac{\rho ^{4}}{r ^{2}}  \right> + \bar{s} ^{zz} (z + R _{\oplus}) ^{2} \left< U \frac{\rho ^{2}}{r ^{2}} \right> \right] \notag \\ & \phantom{=}  - i \frac{ \hbar}{2 m c ^{2} \sigma ^{2}} \bar{s} ^{az} p _{a} (z+ R _{\oplus}) \left< \frac{\rho ^{2}}{r ^{2}} U \right> , \label{V311-FIN}  \\ \left< V _{3} \right> ^{(1,2)} &= \frac{\hbar ^{2}}{4m c ^{2}} \left\lbrace (2+3 \bar{s} ^{00}) \frac{1}{\sigma ^{2}}   \left<  U \frac{\rho ^{2}}{r ^{2}} \right> + \frac{2 i p _{a} \bar{s} ^{az}}{\hbar}  (z+R _{\oplus}) \left[ \left< \frac{U}{r ^{2}} \right>  - \frac{3}{2} \left< U \frac{\rho ^{2}}{r ^{4}} \right> \right] \right. \notag \\ & \hspace{1.5cm} \left. - \frac{\bar{s} ^{aa}}{\sigma ^{2}}  \left[ \left< U \frac{ \rho ^{2}}{r ^{2}} \right> - \frac{3}{2}  \left< U  \frac{\rho ^{4}}{r ^{4}} \right> \right] + \frac{3}{\sigma ^{2}} \bar{s} ^{zz} (z+R _{\oplus}) ^{2}  \left< U \frac{\rho ^{2}}{r ^{4}} \right>  \right\rbrace ,   \label{V312-FIN} \\ \left< V _{3} \right> ^{(1,3)} &= - \frac{\hbar ^{2}}{16 m c ^{2}} \left[ (2 + 3 \bar{s} ^{00}) \left< \frac{2}{r ^{2}} U - \frac{3 \rho ^{2}}{r ^{4}} U \right> + \bar{s} ^{aa}  \left< - \frac{2}{r ^{2}} U + 9 \frac{\rho ^{2}}{r ^{4}} U - 15  \frac{\rho ^{4}}{2r ^{6}} U  \right> \right. \notag \\ & \phantom{=} \left. + 3 \bar{s} ^{zz} (z + R _{\oplus}) ^{2}  \left< \frac{2}{r ^{4}} U - \frac{5 \rho ^{2}}{r ^{6}} U \right> \right] . \label{V313-FIN}
\end{align}
\end{subequations}
Collecting these terms we find an imaginary function whose coefficient is exactly the same appearing in Eq. (\ref{V2-6}) and therefore it vanishes. The remaining terms can be simplified by computing the required integrals. The final result is
\begin{align}
\left< V _{3} \right> ^{(1)} = \frac{1}{2 (m c ) ^{2} } \left( p _{a} p _{a} + \frac{ \hbar ^{2}}{\sigma ^{2}}  \right) \left< V _{1} \right> , \label{V31-FIN}
\end{align}
where $\left< V _{1} \right>$ is given by Eq. (\ref{V1-FIN}). The second term, $\left< V _{3} \right> ^{(2)}$, simplifies to $\left< V _{3} \right> ^{(2)} = - \frac{1}{4m} \left[ \left< h _{00} \right> \hat{p} ^{2} _{z} - \frac{1}{4} \left< h _{00,zz} \right> \right]$, where we have used the result of Eq. (\ref{def}). Since we are considering contributions up to second order in $z$, the latter term will produce only a constant which does not affect the energy levels of the system. Therefore we are left with
\begin{align}
\left< V _{3} \right> ^{(2)} = \frac{1}{2 (mc) ^{2}} \left< V _{1} \right> \hat{p} ^{2} _{z} . \label{V32-FIN}
\end{align}
Finally, using the function $\left< V _{1} \right>$ in Eq. (\ref{V1-FIN}) we compute (\ref{V31-FIN})+(\ref{V32-FIN}) to obtain Eq. (\ref{V3-FIN}).

Finally, we focus on the term $\langle V _{4} \rangle$, given by Eq. (\ref{V4Pot}). Using the  static gauge conditions (\ref{GaugeCond}) to simplify the calculations, $\langle V _{4} \rangle$ takes the form
\begin{align}
\left< V _{4} \right> = - \frac{1}{2m} \left[ \left< h _{jk} \hat{p} ^{j} \hat{p} ^{k} \right> +  2 (1- \bar{s} ^{00}) \left< U _{,k}  \hat{p} ^{k} \right> + \frac{1}{2}  (1 - \bar{s} ^{00}) \left< U _{,kk} \right> \right] . \label{V4-2}
\end{align}
Let's study each term independently. Using the result of Eq. (\ref{def}), the middle term of the above equation, $\left<  V _{4} \right>  ^{(2)} = - \frac{1}{m} (1- \bar{s} ^{00}) \left< U _{,k}  \hat{p} ^{k} \right>$, can be written as $\left<  V _{4} \right>  ^{(2)} = - \frac{1}{m} (1- \bar{s} ^{00})  \left<  U _{,a}  \hat{p} ^{a} - \frac{1}{2} U _{,zz} \right>$. Computing the expectation value of $U _{,k}  \hat{p} ^{k}$ and using some simple identities we find
\begin{align}
\left<  V _{4} \right>  ^{(2)} &= \frac{\hbar ^{2} }{m} (1- \bar{s} ^{00})  \left< \frac{\rho ^{2}}{\sigma ^{2} r ^{2}} U - \frac{U}{r ^{2}} + \frac{3}{2}  \frac{\rho ^{2}}{r ^{4}} U \right> , \label{V42-3}
\end{align}
which, as we already proved, is zero. The latter term in Eq. (\ref{V4-2}), $\left<  V _{4} \right>  ^{(1)} = - \frac{1}{4m} (1- \bar{s} ^{00}) \left< U _{,kk} \right> $, can be simplified by using the result $U _{,kk} = 4 \pi \hbar ^{2} G M _{\oplus} \delta (\vec{r} _{\perp}) \delta (z+R _{\oplus})$, which follows from the gauge conditions. Therefore we obtain $\left< V _{4} \right>  ^{(3)} = - (\hbar ^{2} / m ) (G M _{\oplus} / \sigma ^{2}) (1- \bar{s} ^{00}) \delta (z+R _{\oplus})$. Since the wave function $\chi _{n} (z)$ is defined along the positive $z$-axis, i.e. $z \in [ 0 , \infty)$, then $\chi _{n} (z) \delta (z+R _{\oplus}) = 0$, and therefore $\left< V _{4} \right>  ^{(3)}$ does not contribute to the energy shifts. This means we are left with $\left< V _{4} \right> ^{(1)} = - \frac{1}{2m} \left< h _{jk} \hat{p} ^{j} \hat{p} ^{k} \right>$, which we conveniently splits as $\left<  V _{4} \right> ^{(1)} = \left<  V _{4} \right> ^{(1,1)} + \left<  V _{4} \right> ^{(1,2)} + \left<  V _{4} \right> ^{(1,3)}$, where 
\begin{subequations}
\begin{align}
\left<  V _{4} \right> ^{(1,1)} &= - \frac{1}{2m c ^{2}} \left<  (2 - \bar{s} ^{00}) U + \bar{s} ^{ij} U ^{ij} + 2 \bar{s} ^{00} U ^{zz} - 2 \bar{s} ^{zi} U ^{zi} \right> \hat{p} ^{2} _{z} , \label{V411}  \\ \left<  V _{4} \right> ^{(1,2)} &= - \frac{1}{2m c ^{2}} \left< \left(  4 \bar{s} ^{00} U ^{zb} - 2 \bar{s} ^{zi}U ^{ib} - 2 \bar{s} ^{bi}U ^{iz} \right) \hat{p} _{b} \right> \hat{p} _{z} , \label{V412} \\ \left<  V _{4} \right> ^{(1,3)} &= - \frac{1}{2m c ^{2}} \left<  \left[ (2 - \bar{s} ^{00}) U + \bar{s} ^{ij} U ^{ij} \right] \hat{p} _{a} \hat{p} _{a} + 2 \left( \bar{s} ^{00} U ^{ab} - \bar{s} ^{ai}U ^{bi} \right) \hat{p} _{a} \hat{p} _{b} \right>  . \label{V413}
\end{align}
\end{subequations}
Computing the involved expectation values, the above equations can be written in the form
\begin{subequations}
\begin{align}
\left<  V _{4} \right> ^{(1,1)} &= - \frac{1}{2m c ^{2}} \left\lbrace  (2 + \bar{s} ^{aa}) \left< U \right> - \frac{1}{2} \left( \bar{s} ^{aa} + 2 \bar{s} ^{00} \right) \left<U \frac{\rho ^{2}}{r ^{2}} \right>  \right\rbrace \hat{p} ^{2} _{z} ,  \label{V411-FIN} \\ \left<  V _{4} \right> ^{(1,2)} &= - \frac{1}{2m c ^{2}}  \left\lbrace \left( 3 \bar{s} ^{00} - \bar{s} ^{zz} \right) i \hbar (z+R _{\oplus}) \left<  U \frac{ \rho ^{2}}{\sigma ^{2} r ^{2}}  \right> - p _{a} \bar{s} ^{za} \left( 2 \left< U \right> -  \left< U \frac{\rho ^{2}}{r ^{2}} \right> \right)  \right\rbrace \hat{p} _{z} , \label{V412-FIN} \\ \left<  V _{4} \right> ^{(1,3)} &= - \frac{1}{2mc ^{2}} \left\lbrace  - \frac{ i \hbar p _{a} \bar{s} ^{az}}{\sigma ^{2}} (z + R _{\oplus})  \left< U \frac{\rho ^{2}}{r ^{2}} \right> + \left( p _{a} p _{a} + \frac{2 \hbar ^{2}}{\sigma ^{2}} \right) \left[ (2 - \bar{s} ^{cc}) \left< U \right> + \frac{3}{2} \bar{s} ^{cc} \left< U \frac{\rho ^{2}}{r ^{2}} \right> \right]  \right. \notag  \\ & \phantom{=} \; \left. - \bar{s} ^{ab} \left( p _{a} p _{b} + \frac{\delta _{ab} \hbar ^{2}}{\sigma ^{2}} \right)  \left< U \frac{\rho ^{2}}{r ^{2}} \right>  + \frac{\hbar ^{2}}{\sigma ^{4}} \left[ (\bar{s} ^{cc} - 2) \left< U \rho ^{2} \right> - \frac{1}{2} (3 \bar{s} ^{00} - \bar{s} ^{zz}) \left< U \frac{\rho ^{4}}{r ^{2}} \right> \right] \right\rbrace .  \label{V413-FIN}
\end{align}
\end{subequations}
Now we have to collect all these terms. After simple algebraic simplifications, one can further verify that the imaginary functions appearing with the coefficients $p _{a} \bar{s} ^{za}$ and $3\bar{s} ^{00} - \bar{s} ^{zz}$ can be written in the form of Eq. (\ref{def}), thus implying that they do not contribute to the energy shifts. Therefore we can disregard such terms. The remaining terms can be worked out in a simple fashion. The final result is given by Eq. (\ref{V41-FIN}).

\end{document}